\documentclass[aps,amssymb,prb,twocolumn,showpacs]{revtex4}
\usepackage{graphicx}
\usepackage{epsfig,amsmath}
\begin{document}

\title{Thermal conductance in a spin-boson model: Cotunneling and low temperature properties}
\author{Tomi Ruokola$^1$}
\email[Correspondence to ]{tomi.ruokola@tkk.fi}
\author{Teemu Ojanen$^2$}
\affiliation{$^1$ Department of Applied
Physics,  Aalto University, P.~O.~Box 11100,
FI-00076 Aalto, Finland}
\affiliation{$^2$ Low Temperature Laboratory, Aalto University, P.~O.~Box 15100,
FI-00076 Aalto, Finland }
\date{\today}
\begin{abstract}
Bosonic thermal transport through a two-level system is analyzed at temperatures below and comparable to the two-level energy splitting. It is shown that in the low-temperature regime transport is dominated by correlated two-boson processes analogous to electron cotunneling in quantum dots under Coulomb blockade. We present a detailed analysis of the sequential-cotunneling crossover and obtain essentially an analytic description of the transport problem. Perturbative analysis is complemented by employing scaling properties of the Ohmic spin-boson model, allowing us to extract an anomalous low temperature scaling of thermal conductance.
\end{abstract}
 \bigskip
\maketitle

\section{introduction}
The spin-boson model describes the interaction between bosonic modes and a two-level quantum system.\cite{leggett} It has been employed in a wide variety of phenomena exhibiting dissipation and loss of quantum coherence. Lately the model has found applications in a nonequilibrium setting where two bosonic reservoirs with different temperatures couple through the two-level system (TLS).\cite{segal,segal2,segal3,segal4,segal5,vel1,vel2,hanggi} These considerations are motivated by efforts to understand thermal properties of artificial molecular and nanoscale structures. The spin-boson model is applicable to a generic situation where bosonic fluctuations couple through a nonlinear region at low enough temperatures so that a truncation to the lowest two energy levels is sufficient. The spin-boson model can be regarded as a bosonic counterpart of the Anderson model which describes dynamics of a single spin-degenerate electron orbital coupled to leads. Although these model systems provide a highly simplified picture of real physical systems, they often capture essential properties successfully.

\begin{figure}[h]
\centering
\includegraphics[width=0.67\columnwidth,clip]{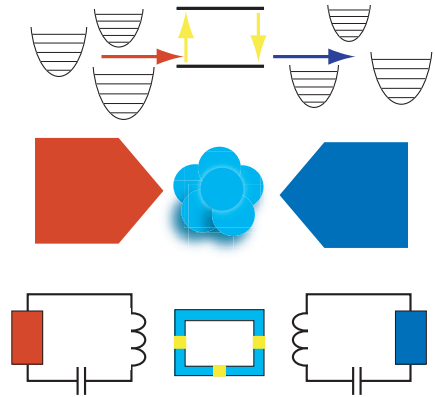}
\caption{(Top) Schematic picture of thermal transport in a two-reservoir spin-boson model.
A two-level system mediates energy transfer between two harmonic oscillator baths held
at different temperatures. Possible experimental realizations: (center) phonons in two bulk leads couple through a nonlinear molecular junction, (bottom) electromagnetic fluctuations of two linear circuits couple through a superconducting qubit.
}\label{scheme}
\end{figure}

The spin-boson model is interesting from the point of view of {\it heattronics}, the manipulation and control of heat in artificial nanostructures.\cite{cahill,carey} Nonlinear elements, a two-level system being the simplest, with tunable parameters are basic building blocks for manipulation of heat flow. Possible realizations of the spin-boson thermal transport are, for example, molecular junctions coupling two phonon reservoirs or electric circuits coupled through a superconducting qubit,\cite{niskanen} see Fig.~\ref{scheme}. In the latter example the qubit can be manipulated by external fields allowing a flexible control of its properties and enabling the investigation of different transport regimes.

In this paper we present a perturbative theory of thermal transport in
the spin-boson model for arbitrary bath temperatures. At temperatures
above or comparable to the TLS energy splitting the reservoirs can excite the
system and transport is dominated by incoherent sequential emission
and absorbtion processes.
%In this regime the heat transport problem
%can be treated adequately in the lowest order approximation in the
%bath coupling.
Except for one numerical simulation,\cite{vel1}
all previous investigations of spin-boson heat
transport\cite{segal,segal2,segal3,segal4,segal5,vel2,hanggi} have
considered only this regime. But when the bath temperatures decrease
below the two-level splitting it is no longer possible to excite the
mediating system and sequential processes are blocked. In this case
one needs to take into account coherent two-boson processes which
correspond to virtual excitation of the two-level system. The
qualitative picture is the same as in electron transport in quantum
dots under Coulomb blockade. At low enough bias voltage the charging
energy blocks exponentially the sequential tunneling processes where
individual electrons hop in and out of the dot
incoherently.\cite{beenakker} The dominant transport processes then
consist of coherent two-electron cotunneling processes where the
population of the dot remains the same after the process.\cite{averin}
In analogy to electron transport, we use here the word ``cotunneling''
for the virtual two-boson processes giving rise to the low temperature
heat conductance.  By calculating two-boson cotunneling rates we
obtain an analytic description for thermal transport below the
sequential-cotunneling crossover. Our results are valid for weak bath
coupling or for an arbitrary coupling strength when the noncommuting
part of the two-level Hamiltonian and coupling operators is small. We
present a detailed analysis of the crossover region and discuss the
applicability of the sequential tunneling approximation. By employing
the scaling equations of the Ohmic spin-boson model we can take into
account renormalization of the spin-boson parameters. This allows us
to complement the perturbative results by scaling effects and extract
an anomalous temperature scaling of cotunneling thermal conductance.

The paper is organized in the following way. In Section \ref{model} we introduce the studied model in detail.  In Section \ref{current} we calculate the cotunneling transition rates of relevant two-boson processes. We present a general formula for heat current from which one can obtain an analytic expressions for the low-temperature and high-temperature limits and study the crossover behavior. In Section \ref{scaling} we use the scaling equations of the Ohmic spin-boson model to calculate temperature scaling of the cotunneling thermal conductance and in Section \ref{conc} we conclude our analysis.

\section{Model}\label{model}
We model our system with the usual spin-boson Hamiltonian, separating the
environmental oscillators into left ($L$) and right ($R$) baths:
\begin{eqnarray}
H&=&H_0 + H_T,\label{H}\\
H_0&=&{\textstyle\frac1 2}\epsilon\sigma_z + {\textstyle\frac1 2}\Delta\sigma_x
+\sum_{i\in L,R}\omega_i(a_i^\dagger a_i + {\textstyle\frac1 2}),\\
 H_T&=&
\sigma_z\sum_{i\in L,R} C_i(a_i + a_i^\dagger).\label{HT}
\end{eqnarray}
The TLS is described by level detuning $\epsilon$ and
transition amplitude $\Delta$, while the baths are characterized
by their spectral functions $\chi_{K}(\omega) = 2\pi\sum_{i\in K}C_i^2
\delta(\omega-\omega_i)$, $K=L,R$.
Equations (\ref{H})-(\ref{HT}) constitute the most generic two-bath
spin-boson system.\cite{leggett}

It is convenient to diagonalize also the TLS part of $H_0$
with a rotation of the pseudospin by an angle $\theta$
in the $x$-$z$ plane, with $\tan\theta = \Delta/\epsilon$.
In this new
basis we have
\begin{eqnarray}
H_0 &=& {\textstyle\frac1 2}\omega_0\sigma_z
+\sum_{i\in L,R}\omega_i(a_i^\dagger a_i + {\textstyle\frac1 2}),\\
H_T &=&
(\sigma_z\cos\theta\ - \sigma_x\sin\theta)
\sum_{i\in L,R} C_i(a_i + a_i^\dagger),
\end{eqnarray}
where $\omega_0 = \sqrt{\epsilon^2 + \Delta^2}$.
It should be noted that the validity of
our perturbation expansion requires that the TLS level broadenings
$\Gamma_0$ and $\Gamma_1$, given by Eqs.~(\ref{g0}) and (\ref{g1})
below, are small compared to the level separation $\omega_0$.  Since
$\Gamma_{0,1}$ contain the couplings $C_i$ as well as $\sin\theta$,
our results are valid either for weak coupling or when the
noncommuting part of the coupling is small ($\theta\ll 1$).

\section{Formula for the current}\label{current}
Our calculation of heat current between the baths is based on the
generalized Fermi Golden Rule.\cite{bruus}
Given two eigenstates $|i\rangle$
and $|f\rangle$ of $H_0$, the perturbation $H_T$ induces transitions
between them at a rate
\begin{equation}\label{ifrate}
\Gamma_{i\to f} = 2\pi\big|\langle f|T|i\rangle\big|^2\delta(E_i-E_f),
\end{equation}
where $E_i$ and $E_f$ are energies of the two states, and the scattering
operator $T$ is
\begin{equation}\label{T}
T = H_T + H_TG_0H_T+ H_TG_0H_TG_0H_T+ \dots
\end{equation}
Here $G_0$ is the retarded propagator for $H_0$ at energy $E_i$,
\begin{equation}
G_0 = \frac{1}{E_i-H_0+i\eta}.
\end{equation}
We will consider only those processes
where one quantum of energy is transported between the baths, i.e.,
the initial and final states are related as
\begin{equation}\label{final}
|f\rangle =
\frac{1}{\sqrt{(N_{K^\prime}+1)N_K}}\,a_{K^\prime}^\dagger a_K|i\rangle,
\end{equation}
where $K$ and $K^\prime$ refer to arbitrary oscillators
in different baths and $N_i$ are their occupation numbers.
In principle the initial and final states of the TLS could be
different but below we show that these inelastic processes
vanish for two-boson transport.
Restriction to
Eq.~(\ref{final}) forbids processes with an odd number of bosons
and therefore we only keep the even powers of $H_T$
in Eq.~(\ref{T}), yielding\cite{flensberg}
\begin{equation}
T = H_TGH_T,
\end{equation}
where $G$ is solved from a Dyson equation as
\begin{equation}
G = \frac{1}{E_i-H_0-\Sigma},
\end{equation}
with self-energy $\Sigma = H_TG_0H_T$.
We consider the self-energy to the lowest order
and keep only the imaginary part which regularizes
the divergence at $\omega_0$. We will now show that in this
case the off-diagonal elements of $\Sigma$ can be neglected.
Let us decompose the $T$ matrix elements as
\begin{equation}\label{tdec}
\langle f|T|i\rangle = \sum_{j,k,\sigma,\sigma^\prime}
\langle f|H_T|k\sigma^\prime\rangle
\langle k\sigma^\prime|G|j\sigma \rangle
\langle j\sigma|H_T|i \rangle,
\end{equation}
where $|j\rangle$ and $|k\rangle$ are states in the bath subspace,
and $|\sigma\rangle$ and $|\sigma^\prime\rangle$ in the TLS subspace.
Similarly we can write $\Sigma$ as
\begin{equation}\label{sigdec}
\begin{split}
\langle k\sigma^\prime|\Sigma|j\sigma \rangle
=
\sum_m\bigg\{
&\frac{\langle k\sigma^\prime|H_T|m\sigma \rangle
\langle m\sigma|H_T|j\sigma \rangle}
{E_i-E_{m\sigma}+i\eta}
+\\
+&
\frac{\langle k\sigma^\prime|H_T|m\bar{\sigma} \rangle
\langle m\bar{\sigma}|H_T|j\sigma \rangle}
{E_i-E_{m\bar{\sigma}}+i\eta}
\bigg\},
\end{split}
\end{equation}
where $|\bar{\sigma}\rangle$ is the state with the spin flipped
with respect to $|\sigma\rangle$. As noted above, we will only be
concerned with the imaginary part of the self-energy, and for the
first term on the RHS it is nonzero when $E_i=E_{m\sigma}$.
On the other hand, to lowest order the self-energy is used to regularize
the divergence of sequential energy transfer, and in that case the
coupling terms $H_T$ in Eq.~(\ref{tdec}) produce energy-conserving
processes, i.e., $E_f=E_{k\sigma^\prime}$ and $E_i=E_{j\sigma}$.
Thus we have $E_{m\sigma} = E_{j\sigma}$, implying $|m\rangle = |j\rangle$
and $\langle m\sigma|H_T|j\sigma \rangle=0$. The first term of
Eq.~(\ref{sigdec}) can therefore
be neglected. For the second term a non-vanishing imaginary part requires
$E_i=E_{m\bar{\sigma}}$, Eq.~(\ref{ifrate}) gives $E_i=E_f$, and the sequential
condition is $E_f=E_{k\sigma^\prime}$, which combine to $E_{m\bar{\sigma}} = E_{k\sigma^\prime}$.
Therefore if $\sigma^\prime = \bar{\sigma}$ we have $|m\rangle = |k\rangle$ and the second
term also vanishes. Only the spin-diagonal elements $\sigma^\prime = \sigma$
of the second term give a finite
imaginary part. This also means that within our weak-coupling approximation
only spin-flipping processes, proportional to
$\sigma_x$ in $H_T$, contribute to $\Sigma$.

Next we argue that the self-energy elements which are
off-diagonal in the bath subspace can also be neglected.
From Eq.~(\ref{sigdec}) one can see that
nonvanishing off-diagonal elements must have $|k\rangle = a_1^{(\dagger)}a_2^{(\dagger)}|j\rangle$,
where $a_i^{(\dagger)}$ denotes either a creation or destruction operator for the baths.
In this case the sum contains only two terms, with $|m\rangle = a_1^{(\dagger)}|j\rangle$
or $|m\rangle = a_2^{(\dagger)}|j\rangle$. In contrast,
 for diagonal elements $|j\rangle=|k\rangle$ the sum
goes over all states of the form $|m\rangle = a_m^{(\dagger)}|j\rangle$.
Since each term in the sum is proportional to the coupling, at least
in our weak-coupling approximation the diagonal elements dominate
and the off-diagonal ones can be discarded.
Thus we obtain the expression
\begin{equation}\begin{split}
-{\rm Im}\, \langle j\sigma|\Sigma|j\sigma\rangle &=
\pi\sum_m \big|\langle m\bar{\sigma}|H_T|j\sigma\rangle\big|^2
\delta(E_{j\sigma}-E_{m\bar{\sigma}})\\
&\equiv {\textstyle\frac1 2}\Gamma_{j\sigma},
\end{split}
\end{equation}
that is, the imaginary part of the self-energy element
is half of the Golden-Rule transition
rate out of state $|j\sigma\rangle$. Collecting everything yields
\begin{equation}\label{ifrate2}
\Gamma_{i\to f} = 2\pi\bigg|\sum_{j,\sigma}\frac{\langle f|H_T|j\sigma\rangle
\langle j\sigma|H_T|i\rangle}{E_i-E_{j\sigma}+i{\textstyle\frac 1 2}\Gamma_{j\sigma}}
\bigg|^2\delta(E_f-E_i).
\end{equation}
Evaluating the sum over the intermediate states $|j\sigma\rangle$ is straightforward.
We note that the calculation shows that even for cotunneling (with $E_i\ne E_{j\sigma}$)
it is again only the spin-flipping terms, proportional to $\sigma_x$ in $H_T$, which
give a non-vanishing contribution. The coupling term proportional to $\sigma_z$ is therefore
completely irrelevant for the two-boson weak-coupling current. This is intuitively
plausible since such a term only produces a small shift in the system energy levels.

Next we replace the lifetime $\Gamma_{j\sigma}^{-1}$ in Eq.~(\ref{ifrate2})
with the lifetime of a state
where the TLS subspace is in state $|\sigma\rangle$
but the bath subspaces are in thermal states. Then
$\Gamma_{j\sigma}$ becomes independent of the bath state $|j\rangle$,
and we call the two possible values $\Gamma_0$ and
$\Gamma_1$, corresponding to ground state
and excited TLS, respectively. A short calculation gives
\begin{eqnarray}
\Gamma_0 &=& \Gamma_L e^{-\omega_0/T_L} + \Gamma_R e^{-\omega_0/T_R},\label{g0}\\
\Gamma_1 &=& \Gamma_L + \Gamma_R,\label{g1}\\
\Gamma_K &=& \sin^2\theta\,\chi_K(\omega_0)
\big[n_K(\omega_0) + 1\big],\ K=L,R.
\end{eqnarray}
Here $\Gamma_K$ is the decay rate of the excited TLS into bath $K$
with temperature $T_K$ and corresponding Bose distribution $n_K$.

To derive a formula for the heat current we first write down an expression
for heat absorbed from the left bath and emitted to the right bath, with $\sigma$
being the initial (and final) state of the TLS
\begin{equation}\label{jlr}
J^{(\sigma)}_{L\to R} = \sum_{i,f} \omega_{f_L} \Gamma_{i\to f} W_i,
\end{equation}
where the sum is over all possible initial states of the reservoirs
(with a thermal distribution $W_i$), and all final states of the form
(\ref{final}) with $K=f_L$ an arbitrary oscillator
in the left bath and $K^\prime=f_R$ in the right bath.
Energy conservation requires that $\omega_{f_L} = \omega_{f_R}$.
Current in the other direction, $J^{(\sigma)}_{R\to L}$,
is defined analogously, and the net heat flow is
$J^{(\sigma)} = J^{(\sigma)}_{L\to R} - J^{(\sigma)}_{R\to L}$
which can be evaluated as
\begin{eqnarray}
J^{(\sigma)} &=& \frac{\sin^4\theta}{2\pi}\int_0^\infty d\omega
\,\omega\chi_L(\omega)\chi_R(\omega)\big[n_L(\omega)-n_R(\omega)\big]\nonumber\\
&&\times\bigg|
\frac{1}{\omega-\omega_0\pm\frac i 2\Gamma_{\bar{\sigma}}} -
\frac{1}{\omega+\omega_0\mp\frac i 2\Gamma_{\bar{\sigma}}}
\bigg|^2,\label{jm}
\end{eqnarray}
where upper signs refer to $\sigma=0$,
lower signs to $\sigma=1$.

\begin{figure}[ht]
\centering
\includegraphics[width=0.99\columnwidth,clip]{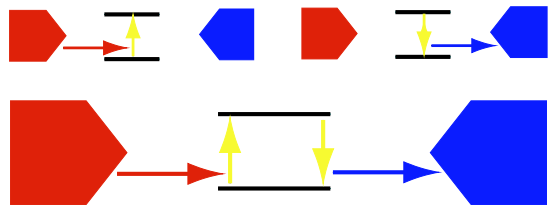}
\caption{(Top) At temperatures comparable to the two-level energy splitting transport is dominated by incoherent sequential processes. (Bottom) At low temperatures it is no longer possible to excite the two-level system and transport is dominated by coherent two-boson processes creating virtual excitations.}\label{cotunnelling}
\end{figure}
\begin{figure}[h]
\centering
\includegraphics[width=0.8\columnwidth,clip]{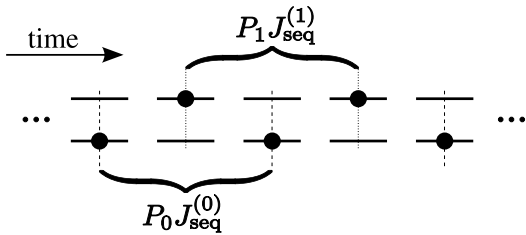}
\caption{Evolution of the TLS state due to sequential tunneling events.
The current carried by these tunneling events can be calculated in two different
ways. On one hand, we can start with
the TLS in ground state, and then one current-carrying process
flips the state twice, and the next process again starts
from the ground state. The average current transported by these
processes is $J_{\rm seq}=P_0J^{(0)}_{\rm seq}$. On the other
hand, we can describe the same sequence of events by starting with
the TLS in the excited state, giving $J_{\rm seq}=P_1J^{(1)}_{\rm seq}$.
Thus the two expressions for sequential current are not additive but
alternatives to one another.
}
\label{partiti}
\end{figure}

We still have to combine the two partial currents $J^{(\sigma)}$ to obtain the
full heat flow through the system.
First note that under steady-state conditions the probability for the TLS to
be in ground state is $P_0 = \Gamma_1/(\Gamma_0+\Gamma_1)$, and
for the excited state $P_1=1-P_0$.
Let us then split the currents into sequential
and cotunneling contributions,\cite{nasse} $J^{(\sigma)} =
J^{(\sigma)}_{\rm seq} + J^{(\sigma)}_{\rm cot}$,
see also Fig.~\ref{cotunnelling}.
For sequential processes the intermediate state has the 
TLS spin flipped, for cotunneling processes this flipping is only virtual
and does not contribute to $P_{0/1}$.
Now, for sequential transport, one current-carrying process consists of two
incoherent tunneling events, and if we start with, say,  the TLS in ground state,
then the intermediate state is excited. But this intermediate state
can equally well be seen as the initial state of another current-carrying process.
Because of this overlap, one must be careful to avoid double counting,
and as further elucidated in
Fig.~\ref{partiti}, the full sequential current is $J_{\rm seq}=
P_0J^{(0)}_{\rm seq} = P_1J^{(1)}_{\rm seq}$.

On the other hand, cotunneling events are non-overlapping and therefore
they contribute additively, $J_{\rm cot}=
P_0J^{(0)}_{\rm cot} + P_1J^{(1)}_{\rm cot}$.
Total heat current through the system, $J=J_{\rm seq} + J_{\rm cot}$, is therefore
\begin{equation}\label{jlong}
J = P_0J^{(0)} + P_1J^{(1)} - J_{\rm seq}.
\end{equation}
The sequential current can be calculated
by taking the difference of the forward and backward Golden Rule rates
at either junction, e.g., for the left junction,
\begin{equation}\label{seqdef}
J_{\rm seq} = \omega_0(P_0\Gamma_L e^{-\omega_0/T_L}-P_1\Gamma_L).
\end{equation}
We can still simplify Eq.~(\ref{jlong}). As will be shown below,
at high temperatures sequential tunneling dominates and all three
terms on the RHS of Eq.~(\ref{jlong}) are equal in magnitude
and thus the last two terms cancel. They can also be neglected
at low temperatures since then $P_1$ and $J_{\rm seq}$
are exponentially suppressed. For these limits we therefore have
\begin{equation}\label{fullj}
J = P_0J^{(0)}.
\end{equation}
Numerical calculations suggest that this expression is
actually
valid within a few precent at all temperatures.
Equation of the
same form has been previously used by Flensberg\cite{flensberg}
in the context of electron cotunneling.

\subsection{High-temperature limit}

Next we will derive the high and low-temperature limits for
Eq.~(\ref{fullj}), corresponding to sequential and cotunneling,
respectively. At high temperatures resonant energy transfer
at $\omega=\omega_0$ dominates, and therefore
the second energy denominator in Eq.~(\ref{jm}) can be dropped.
Contribution from the resonance peak can be calculated by
taking the limit $\Gamma_{0,1}\to0$, implying
$|\omega-\omega_0\pm\frac i 2\Gamma|^{-2}\to\frac{2\pi}{\Gamma}
\delta(\omega-\omega_0)$. The current is then
\begin{equation}\label{seq}
J = \frac{\sin^2\theta\, \omega_0\chi_L(\omega_0)\chi_R(\omega_0)
\big[n_L(\omega_0)-n_R(\omega_0)\big]}{
\chi_L(\omega_0)\big[2n_L(\omega_0)+1\big]
+\chi_R(\omega_0)\big[2n_R(\omega_0)+1\big]
}.
\end{equation}
This expression is equivalent to the sequential tunneling current
calculated from Eq.~(\ref{seqdef}), and to the result derived by Segal and Nitzan.\cite{segal}

The processes leading to Eq.~(\ref{seq}) correspond to sequential absorption-emission of bosons as depicted in Fig.~\ref{cotunnelling}. In this limit the heat current is sensitive only to the value of the bath spectral functions at resonance. At temperatures $T_{L,R}\ll\omega_0$ expression (\ref{seq}) becomes exponentially small, signalling the breakdown of the sequential tunneling approximation.

\subsection{Low-temperature limit}
In the low-temperature limit, $T_{L,R}\ll\omega_0$ we can ignore
the $\omega$-dependence of the energy denominators in Eq.~(\ref{jm}).
In addition, the excitation of the TLS is exponentially suppressed, so that
$J=J^{(0)}$. Then the general expression reduces to
\begin{eqnarray}\label{cot}
J = \frac{2\sin^4\theta}{\pi\omega_0^2}\int_0^\infty d\omega\,
\omega\chi_L(\omega)\chi_R(\omega)\big[n_L(\omega)-n_R(\omega)\big].
\end{eqnarray}
Transport is dominated by two-boson cotunneling processes during which
one boson is absorbed and one emitted as in Fig.~\ref{cotunnelling}.

Let us remark that both Eqs.~(\ref{seq}) and (\ref{cot}) can be cast in Landauer
form, $J=\int d\omega\,\omega{\cal T}(\omega)\big[n_L(\omega)-n_R(\omega)\big]$,
where ${\cal T}(\omega)$ is the transmission function. An important difference between
the two cases is that in the sequential regime ${\cal T}$ depends on the bath
temperatures while for cotunneling the transmission is temperature-independent.
One consequence is that the cotunneling current is symmetric with respect
to the reversal of the temperature gradient, and therefore in contrast to the
high-temperature case,\cite{segal} the spin-boson system
cannot act as a thermal rectifier at low temperatures.

For concreteness we calculate the cotunneling current
in the case where the bath spectral densities have a power-law form $\chi_K(\omega)=
A_K\omega^{s_K}$, with $0<s_K<1$ corresponding to sub-Ohmic,
$s_K=1$ to Ohmic, and $s_K>1$ to super-Ohmic environment.
Then we have
\begin{equation}\label{lowt}
J = \frac{2A_LA_R}{\pi\omega_0^2}\sin^4\theta\,
\Gamma(\nu)\zeta(\nu)\,(T_L^\nu-T_R^\nu),
\end{equation}
where $\nu = 2+s_L+s_R$, and $\Gamma$ and $\zeta$ are the
gamma and zeta function, respectively. It is interesting to note that the temperature dependence of the current is sensitive only to the sum of the exponents $s_L+s_R$ and a precise nature of the individual baths is irrelevant.

\subsection{Crossover}

\begin{figure}[h]
\centering
\includegraphics[width=\columnwidth,clip]{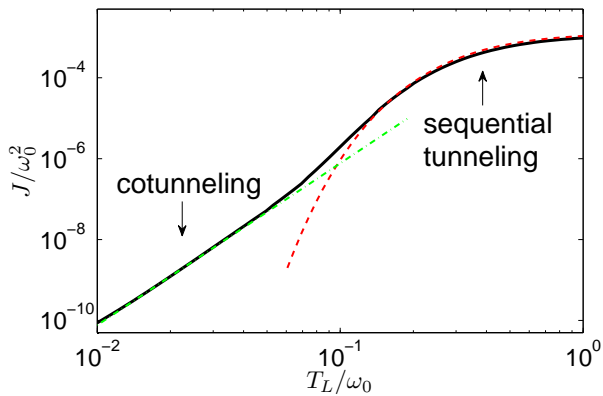}
\caption{Current through the system as a function of left-bath
temperature $T_L$.
Solid line is the numerical solution of Eq.~(\ref{fullj}),
dashed line is the high-temperature approximation from Eq.~(\ref{seq}),
and dash-dotted line is the low-temperature approximation from Eq.~(\ref{lowt}).
Both baths are ohmic, $s_L=s_R=1$, with equal
coupling strengths $A_L=A_R=0.1$, and $\sin\theta=1$. Right-bath
temperature is $T_R=0.95\,T_L$.}\label{jtplot}
\end{figure}

In Fig.~\ref{jtplot} we plot the numerical solution of Eq.~(\ref{fullj})
together with the limiting forms given by Eqs. (\ref{seq}) and (\ref{lowt}).
We see that the crossover between sequential tunneling and cotunneling
is quite sharp, and the two analytical formulas are very well applicable outside
a narrow transition region. By patching together the two limiting expressions one achieves an accurate analytical description of the transport problem at all temperatures.
The crossover temperature depends on the system parameters logarithmically and
it is generally of the order of $0.1\,\omega_0$, below which the naive Golden-Rule treatment becomes invalid.

\section{Low temperature properties
from renormalization group}\label{scaling}

In the linear-response limit, $T_R\to T_L\equiv T$, Eq.~(\ref{lowt})
shows that in the cotunneling regime
thermal conductance $G$ depends on the temperature as
$G\propto T^{1+s_L+s_R}$. This result was derived using perturbation theory and as such relies on an assumption of weak coupling between the TLS an the baths. In principle it is possible to calculate higher orders of perturbation theory and find corrections but this becomes soon unfeasible.
Instead of carrying out a more complete perturbative calculation, in this section
we use a renormalization group (RG) arguments to improve our analysis and find correction to the temperature
exponent of $G(T)$.

To set up the RG framework, we introduce a cutoff frequency $\omega_c$
above which the bath response functions $\chi_{L,R}$ vanish.
The initial value of $\omega_c$ is irrelevant for our purposes.
We then perform the RG step by lowering the cutoff and integrating out
all bath oscillators with frequencies above the new cutoff.
This system is again described by a Hamiltonian of the form (\ref{H})
but with a new set of parameters. The process is continued until $\omega_c$
is equal to the temperature $T$. This way the system parameters,
which we have so far treated as constants, become
effectively temperature dependent and give an extra contribution
to $G(T)$.

\begin{figure}[h]
\centering
\includegraphics[width=\columnwidth,clip]{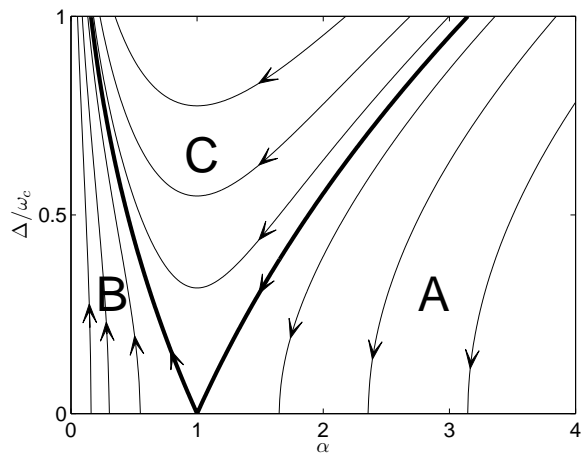}
\caption{Renormalization group flow of the parameters
$\alpha$ and $\Delta/\omega_c$ in the Ohmic spin-boson model, as given by
Eqs.~(\ref{flow1}) and (\ref{flow2}). Two separatrices (thick lines)
divide the flows into three qualitatively different
regimes.}\label{flows}
\end{figure}
The actual RG calculation is performed by the well-known Coulomb gas
mapping.\cite{leggett} This procedure is only available
for Ohmic baths, and therefore for the rest of
this section we set $s_L=s_R=1$. The partition function is transformed to a form which is identical to the anisotropic Kondo model after the Yuval-Anderson mapping.\cite{yuval} The renormalization procedure in this representation was worked out by Anderson, Yuval and Hamann in the seminal paper [\onlinecite{anderson}]. To conform with established notation
we introduce dimensionless coupling constants $\alpha_{L,R}$
such that the bath spectra are given by $\chi_K(\omega)
=\pi\alpha_K\omega$. The RG flow equations for the one-bath spin-boson
model\cite{leggett,costi} are directly applicable for several baths,
\begin{eqnarray}\label{flow1}
\frac{d(\Delta/\omega_c)}{d\ln\omega_c}&=&-(1-\alpha)\bigg(\frac{\Delta}{\omega_c}\bigg)\\
\frac{d\,\alpha}{d\ln\omega_c}&=&\alpha\bigg(\frac{\Delta}{\omega_c}\bigg)^2,
\label{flow2}
\end{eqnarray}
with $\alpha=\alpha_L+\alpha_R$ being a combined coupling for both baths.
These equations are valid for $\alpha<4$ and
$\Delta/\omega_c\lesssim1$. The flow of the parameters with decreasing
$\omega_c$ is depicted in Fig.~\ref{flows}.
There a three qualitatively different regimes which we call A, B, and C.

In regime A the renormalization of $\alpha$ is small in a realistic case where the bare value of $\Delta/\omega_c$ is small, so we can treat $\alpha$ as a constant.
Then solving Eq.~(\ref{flow1}) yields\cite{leggett,costi}
\begin{equation}\label{deltar}
\Delta = \Delta_0\left(\frac{\omega_c}{\omega_{c0}}\right)^\alpha,
\end{equation}
where $\Delta_0$ and $\omega_{c0}$ are the initial, unrenormalized values
of $\Delta$ and $\omega_c$. The cutoff is taken all the way down
to the temperature $T$, and we obtain $\Delta\propto T^\alpha$. Now we can
again examine the $T$ dependence of conductance. From Eq.~(\ref{lowt}),
\begin{equation}\label{reg}
G\propto\frac{\sin^4\theta}{\omega_0^2}T^3 =
\frac{\Delta^4}{(\Delta^2+\epsilon^2)^3}T^3 \propto\left\{
\begin{array}{ll}
T^{3+4\alpha}, & \epsilon\gg\Delta\\
T^{3-2\alpha}, & \epsilon\ll\Delta
\end{array}\right.
\end{equation}
Thus the RG calculation gives a non-perturbative correction
to the bare $T^3$ behavior depending on the interaction strength $\alpha$.
We have extracted two limiting forms
of this correction, depending on the relative magnitudes
of the TLS detuning and the renormalized tunneling element. Since $\Delta$ always renormalizes to zero, in the asymptotic low-temperature limit thermal conductance always obeys $G\propto T^{3+4\alpha}$ even if the bare parameters satisfy $\epsilon\ll\Delta$.

We would like to make a brief comment regarding the validity of the above considerations. There are two possible restrictions, namely the
validity of flow equations, Eqs.~(\ref{flow1}) and (\ref{flow2}),
and the validity of the perturbation calculation for the current,
Eq.~(\ref{lowt}).
Since $\alpha$ and $\Delta/\omega_c$ move towards smaller values,
the flow equations remain valid throughout the process, all the way
to $\omega_c\to 0$. Thus Eq.~(\ref{reg}) can be used even in the
zero temperature limit. On the other hand, validity of Eq.~(\ref{lowt})
requires that the perturbation expansion parameter,
essentially $\alpha\sin^2\theta$, is small. In regime A, where $\alpha>1$,
this means that $\theta$ must be small, i.e., $\Delta/\epsilon\ll1$, so the possibility $G\propto T^{3-2\alpha}$ is not realized in the region of validity of the perturbation theory and must be discarded.

In regime B we can also accurately approximate $\alpha$ with a constant value,
and therefore Eq.~(\ref{reg}) applies also to this case. However,
now the flow takes $\Delta/\omega_c$ towards larger values eventually
leaving the region of validity of the flow equations.
Therefore the flow should be terminated when $\Delta/\omega_c=1$, and at
this point the value of the cutoff is\cite{leggett,costi}
\begin{equation}
\omega_{c,\min}=\omega_{c0}\left(\frac{\Delta_0}{\omega_{c0}}
\right)^\frac{1}{1-\alpha}.
\end{equation}
Thus Eq.~(\ref{reg}) can only be used down to temperatures $T\sim\omega_{c,\min}$.
The perturbation parameter $\alpha\sin^2\theta$ is always small for any $\theta$ by taking a small enough $\alpha$,
so both limiting cases of Eq.~(\ref{reg}) are relevant above the cutoff temperature $\omega_{c,\min}$. In the asymptotic weak-coupling limit $\alpha\to 0$ one recovers the simple $G\propto T^3$-dependence as expected by the perturbation calculation.

Regime C has the most complicated flow and $\alpha$ cannot be treated
as a constant, prohibiting simple analytical estimates. Also in this case $\Delta/\omega_c$ is a relevant perturbation eventually flowing to strong coupling. In principle the flow equations can still be used to analyze the temperature dependence of thermal conductance above the cutoff energy by numerically solving the flow.

\section{Conclusions}\label{conc}

We studied thermal transport in the spin-boson model at temperatures below the TLS energy splitting. At low temperatures a naive Golden-Rule approximation becomes invalid and transport is dominated by two-boson processes. Qualitatively the behavior is similar to electronic transport in quantum dots under Coulomb blockade when lowest-order processes are exponentially suppressed. Applying perturbation methods we obtained an analytical description of low-temperature transport and investigated the crossover between the low-temperature and high-temperature limits. The crossover region is sharp and an accurate analytical description at all temperatures is provided by patching the two limiting cases together. In the case of reservoirs with simple power law spectral densities $\chi_{L/R}(\omega)\propto\omega^{s_{L/R}}$ the low-temperature thermal conductance obeys $G(T)\propto T^{1+s_L+s_R}$. Going beyond perturbation theory, we applied renormalization group arguments to extract anomalous temperature scaling of thermal conductance in the Ohmic case. The power law predicted by simple perturbation theory was modified by interaction-dependent correction which becomes significant outside the weak-coupling regime.

The studied problem is not only important for future applications, but also relevant in systems available for experimental studies presently. One of the most flexible systems to probe our predictions is a superconducting flux qubit coupled to two linear circuits acting as reservoirs.\cite{niskanen, ojanen2} In this realization the electromagnetic fluctuations of the reservoirs play the role of bosonic modes in the spin-boson model. Using external fields to tune the energy splitting of the qubit it is possible to address different regimes in a controlled manner. Considering the generic role of the spin-boson model, it is expected that our results are important for a variety of applications in heattronics of molecular and nanoscale systems in the future.

Upon completion of the manuscript we learned about a new work by Wu and Segal.\cite{bo}
They have studied a system similar to ours and their results agree with Eq.~(\ref{cot}).
\acknowledgments
Authors would like to thank Antti-Pekka Jauho and Matti Laakso for useful discussions. One of the authors (T.O.) would like to thank the Academy of Finland for support.

\end{document}